# A Reflective Approach to Providing Flexibility in Application Distribution

Álvaro J. Rebón Portillo, Scott M. Walker, Graham N. C. Kirby, Alan Dearle

*School of Computer Science, University of St Andrews,*

*St Andrews, Fife, KY16 9SS, Scotland, UK.*

rafda@dcs.st-and.ac.uk    www-systems.dcs.st-and.ac.uk/rafda

**Abstract**

Current middleware systems suffer from drawbacks. Often one is forced to make decisions early in the design process about which classes may participate in inter-machine communication. Further, application level and middleware specific semantics cannot be separated, forcing an unnatural design. The RAFDA project proposes to address these deficiencies by creating an adaptive, reflective framework that enables the transformation of non-distributed applications into semantically equivalent applications whose distribution architecture is flexible. This paper describes the code transformation techniques that have been developed as part of the project. The system enables the distribution of a program according to a flexible configuration without user intervention. Remote and non-remote versions of an object become interchangeable. The distributed program can adapt to its environment by dynamically altering its distribution boundaries.

## 1  Introduction

Current middleware systems [1, 2, 3, 4, 5] suffer from drawbacks. Often one is forced to make decisions early in the design process about which classes may participate in inter-machine communication. Further, application level and middleware specific semantics cannot be separated, forcing an unnatural design.

The Reflective Architecture Framework for Distributed Applications (RAFDA) project addresses these deficiencies by creating an adaptive, reflective framework that enables the transformation of non-distributed applications into semantically equivalent applications (modulo network failure) whose distribution boundaries are flexible.

One approach to providing flexibility of distribution boundaries requires the ability to substitute an object with a proxy to a remote instance. Achieving this requires research in the areas of code transformation, type systems, reflection and distributed systems.

This paper discusses the code transformation techniques that have been developed as part of the project. The system described identifies points of substitutability and extracts an interface for each substitutable class. Every reference to a substitutable class must then be transformed to use the extracted interface. Various proxies implementing the interface for a class provide alternative remote versions, e.g. SOAP-based, RMI-based, CORBA-based, etc. The use of interfaces makes non-remote and remote versions of a class interchangeable. The resulting distributed program can adapt to its environment by dynamically altering its distribution boundaries. Policy dictates which classes are substitutable and which proxy implementations are used.

Figure 1 shows a typical distribution scenario. Objects of class A and class B hold references to a shared instance of class C. The application is transformed so that the instance of C is remote to its reference holders. The local instance of C is replaced with a proxy, Cp, to the remote implementation, C'.

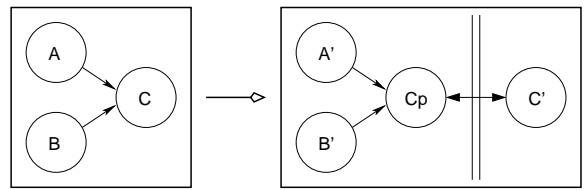

Figure 1: Typical re-distribution scenario.

A reliance on the availability of source code limits the set of applications that can be transformed by a system.

For this reason the transformation process operates at the bytecode level. Java [6, 7] and the Byte Code Engineering Library (BCEL) [8] were chosen for the implementation, but the approach described is not specific to these technologies.

The rest of the paper has been arranged into three sections: Section 2 describes and justifies the transformations performed. It also explains decisions and restrictions resultant from using Java. Section 3 explores related work. Section 4 concludes the paper by describing the extent of the work completed and future work.

## 2 Approach

The overall approach is to extract an interface from each substitutable class in the original application, capturing the functionality of that class. Multiple implementations of the extracted interfaces are then generated in order to provide the various distributed versions of the original classes. The generated code uses only interface types so that substitution of implementations can be made easily, leaving object creation as the only implementation-aware operation. More specifically, for each class A in the original application, the following are generated:

- an interface, A_O_Int, capturing the functionality of A's instance members, together with a set of implementing classes: A_O_Local provides the non-remote version while various proxy classes communicate with remote objects using different protocols, e.g. A_O_Proxy_SOAP would provide an implementation using SOAP as the transport layer;

- an interface, A_C_Int, capturing the functionality of its static members, together with a set of implementing classes, {A_C_Local, A_C_Proxy_RMI, ...};

- an *object factory* class, A_O_Factory, providing the necessary methods for object creation and initialisation; the object creation method contains the policy determining which of the classes implementing A_O_Int will be used; and,

- a *class factory* class, A_C_Factory, providing the methods for class (static members) implementation discovery and initialisation.

Figure 2 shows a sample application class that will be used in the subsequent sections to illustrate the transformations. Although the transformations take place at bytecode level, Java has been used for clarity.

```java
public class X {
    private Y y;
    public X(Y y) { this.y = y; }
    protected int m(long j) { return y.n(j); }
    static final Z z = new Z(Y.K);
    static int p(int i) { return z.q(i); }
}
```

Figure 2: Sample application class X.

### 2.1 Instance Members

This section describes the construction of A_O_Int and its associated implementations. The approach involves the interception of all operations on an object and choosing how each operation should be satisfied using an appropriate implementation. This is not possible for direct field access. The first step of the transformation is therefore to turn every attribute into a property by generating get and set methods for each attribute.

Interfaces are assumed to contain only public members. Since all members will be accessed via an interface they must be made public. This process is safe as the transformations are performed on code that has already been verified by a standard compiler.

A default, parameter-less constructor is added to the class to perform initialisation routines particular to the implementation. All the original constructor functionality is moved to the factories.

Finally, implementations for the methods in the transformed class are provided. Affected type signatures and method calls must be adapted to use the interfaces.

Figure 3 illustrates the result of these transformations for the sample application class X of Figure 2.

### 2.2 Static Members

Because interfaces cannot capture the static functionality of a class, static members are made non-static in order that they can be transformed as described for instance members. The uniqueness semantics of the static members is guaranteed by requiring that all generated implementations be singletons.

Figure 4 illustrates the result of these transformations, again for the sample application class X of Figure 2.

```java
public interface X_O_Int {
  Y_O_Int get_y();
  void set_y(Y_O_Int y);
  int m(long j);
}

public class X_O_Local implements X_O_Int {
  private Y_O_Int y;
  public X_O_Local() { }
  public Y_O_Int get_y() { return y; }
  public void set_y(Y_O_Int y) { this.y = y; }
  // get_y() and n(j) below are interface calls
  public int m(long j) { return get_y().n(j); }
}

public class X_O_Proxy_SOAP implements X_O_Int {
  public X_O_Proxy_SOAP() {
    // SOAP-specific initialisation
  }
  // these methods perform SOAP calls
  // on the real remote object
  public Y_O_Int get_y() { ⋯ }
  public void set_y(Y_O_Int y) { ⋯ }
  public int m(long j) { ⋯ }
}

public class X_O_Proxy_RMI implements X_O_Int {
  ⋮
}
```

Figure 3: Result of instance members transformation for sample class X.

```java
public interface X_C_Int {
  Z_O_Int get_z();
  int p(int i);
}

public class X_C_Local implements X_C_Int {
  private Z_O_Int z;
  public X_C_Local() { }
  public Z_O_Int get_z() { return z; }
  public int p(int i) { return get_z().q(i); }
  // singleton declarations
  private static X_C_Int me = new X_C_Local();
  public static X_C_Int get_me() { return me; }
}

public class X_C_Proxy_RMI implements X_C_Int {
  public X_C_Proxy_RMI() {
    // RMI-specific initialisation
  }
  // these methods perform RMI calls
  // on the real remote object
  public Z_O_Int get_z() { ⋯ }
  public int p(int i) { ⋯ }
}

public class X_C_Proxy_SOAP implements X_C_Int {
  ⋮
}
```

Figure 4: Result of static members transformation for sample class X.

## 2.3 Factories

It is assumed that factory classes are available locally on all participating nodes. It is also assumed that any implementation class required by a factory is available to it.

An object creation method, make, selects which of the implementations is to be used based on some policy.

All access to static members of a class is performed via the singleton implementing these members. A class discovery method, discover, is used to obtain that implementation.

The object creation and class discovery methods are the only potentially implementation-aware methods, and therefore the only ones that could be affected if a new remote implementation is added to or removed from the system.

For every constructor in the original class a matching initialisation method, init, is added to the object factory. The class initialisation method, clinit, matches the static initialiser in the original class. These methods are adapted to take the object or class to be initialised as an extra parameter.

Figure 5 illustrates the result of these transformations, once more for the sample application class X of Figure 2.

## 2.4 Language Specific Issues

There are a number of issues that are not part of the general approach but specific to the use of Java as the target language, in particular, user-defined interfaces, arrays and exception handling. Solutions to all these problems are available but beyond the scope of this paper. Other languages may present different issues but these are representative of the problems of implementing a real system.

```
public class X_O_Factory {
  public static X_O_Int make() { ⋯ }
  public static void init(X_O_Int that, Y_O_Int y) {
    that.set_y(y);
  }
}

public class X_C_Factory {
  public static X_C_Int discover() { ⋯ }
  public static void clinit(X_C_Int that) {
    Z_O_Int t = Z_O_Factory.make();
    Z_O_Factory.init(t, Y_C_Factory.discover().get_K());
    that.set_z(t);
  }
}
```

Figure 5: Factories for sample class X.

It is not practical to inspect or transform code in native methods. Also, some system classes and interfaces have special semantics in the JVM, e.g. to throw an object requires that it extends, directly or indirectly, the java.lang.Throwable class. As a consequence, these special classes and interfaces are not transformed.

A non-transformable class that extends a transformed one would have to inherit from both the instance and static members implementations of its transformed super-class. This would require a multiple inheritance mechanism not available in Java. For this reason, the super-class of a non-transformable class cannot be transformed.

References in a non-transformable class cannot be altered and thus classes and interfaces it refers to should remain available in their original forms. This prevents transformation of classes and interfaces referenced by a non-transformable class.

A class that cannot be transformed cannot be substitutable. About 40% of the 8,200 classes and interfaces in JDK 1.4.1 cannot be transformed. This percentage would increase if the user code contains native methods which refer to a JDK class.

## 3  Related Work

An alternative approach to this problem is to generate wrappers for every class, as opposed to directly transforming code as described here. Wrappers act as proxies to local objects, by encapsulating an object and intercepting all access requests to that object. There is a wrapper per instantiated object and all references to that object are altered to refer to the wrapper.

Although much simpler in terms of implementation, this introduces significantly greater overhead and does not offer solutions to any of the current limitations.

A hybrid of the described approach and the wrapper approach was investigated. This presented problems with dynamic inheritance that made it not feasible as a solution to the problem.

Similar code transformations are used in Orthogonally Persistent Java [9]. The goal of persistence presents a different problem space to that of flexible distribution and object migration but the core approaches are comparable. In the RAFDA project the static component of a class must be handled in a more complex fashion as instances of a class may be spread across multiple address spaces.

ProActive PDC [10] is a Java library that offers dynamic object distribution and migration. It uses bytecode transformation to generate remotely accessible code from that written without distribution in mind. The programmer still must determine statically which objects are to be remotely accessible. The ProActive architecture is similar to the wrapper generation approach that was also investigated.

JavaParty [11] has a similar purpose to the RAFDA project. A new keyword is added to Java identifying remote objects. JavaParty code is preprocessed into standard Java and RMI based code. JavaParty's transformations take place at a source level and, in contrast to RAFDA, the programmer must determine which objects may be remote at design time.

## 4  Conclusions

The transformations described in this paper act on a non-distributed Java program to produce a componentised, semantically equivalent version. This transformed version can be extended while retaining program semantics in order to provide requirements such as distribution or persistence. This approach has been implemented allowing the creation of a local version of the transformed application that executes within a single address space —the first step in creating a fully distributed version.

Changing applications to span address space boundaries may introduce network failure problems. This makes it impossible to guarantee full preservation of the original application semantics. Although potentially a major problem, this is not critical when restricted to a local area network and the behaviour of practical applications will be investigated.

The next stage in our research will see the implementation of adaptable distribution in the deployed application. A series of techniques for enabling communication between remote objects has been devised and is to be integrated with the transformation techniques to produce a complete mechanism for dynamic distribution reconfiguration. In the longer term it is hoped to develop a complete system for deciding and capturing distribution policy.